# Quiet Sun Magnetism from ViSP Data Multiline Inversions Around 630.1 nm

J. C. Trelles Arjona[1,2,3]  
[1] Laboratory for Atmospheric and Space Physics (LASP), University of Colorado, 1234 Innovation Drive, Boulder, CO 80303, USA; juan.trelles@iac.es  
[2] Instituto de Astrofísica de Canarias (IAC), Vía Láctea s/n, 38205 San Cristóbal de La Laguna, Tenerife, Spain  
[3] Dept. Astrofísica, Universidad de La Laguna, 38205 San Cristóbal de La Laguna, Tenerife, Spain

## Abstract

Quiet Sun magnetism plays an important role in the global energy balance of the solar atmosphere. The study of the magnetism of the quiet Sun has been a major effort in the last decades, and, as a result, very important advances in our knowledge have been achieved. The recent commissioning of the Daniel K. Inouye Solar Telescope, the largest ground-based solar telescope in the world today, provides us with data of high spectropolarimetric quality and high spatial resolution. The combination of data acquired with the ViSP instrument on the DKIST telescope around 630.1 nm and multiline inversions enables the achievement of exceptionally high spatial resolution in the solar atmosphere, both at the surface and in depth. In this paper we determine the $\log(gf)$ values of the spectral lines around 630.1 nm observed with the ViSP instrument and search for the best multi-inversion strategy, by means of MHD simulations, to analyze quiet Sun magnetism data within such spectral range. We present results of quiet Sun magnetism with special emphasis on its evolution with optical depth. In the internetwork, we find the decrease with height of the averaged magnetic field strength and the zone from which the fields become more horizontal. Additionally, we explore the depth-dependent evolution of the network's canopy, uncovering intriguing insights into its thermodynamic and magnetic properties. Notably, we detect a temperature enhancement in very localized areas of the vicinity of the network and distinguish mass motions with opposing velocities inside the flux tubes.

*Unified Astronomy Thesaurus concepts:* Quiet sun (1322); Solar physics (1476); Solar magnetic fields (1503); Spectropolarimetry (1973)

## 1. Introduction

The quiet Sun is the area of the solar atmosphere away from the active regions where the apparent serene granulation is observed. Despite its seemingly peaceful appearance, the quiet Sun harbors a dynamic and complex magnetic environment that plays a crucial role in the overall energetic ensemble of the solar atmosphere (J. Trujillo Bueno et al. 2004). The magnetic fields of the quiet Sun are historically organized according to their strength, structure, and location into two parts. The network magnetic fields are located at the edges of supergranular cells, forming kG structures vertical to the solar surface. The internetwork is formed by disorganized hG fields covering the interior of supergranular cells. In recent decades, significant advances have been made in understanding the magnetic properties of the quiet Sun. These studies have revealed intricate details about the spatial distribution of magnetic fields, the thermodynamic structure of the solar atmosphere, and the dynamics of plasma motions (see L. R. Bellot Rubio & D. Orozco Suárez 2019, and references therein). Nevertheless, some questions remain under debate today and have not yet been studied in depth.

A topic lacking a definitive consensus on quiet Sun magnetism is the variation with height of the weak fields in the internetwork. J. O. Stenflo (2013), utilizing the ZIMPOL instrument (A. M. Gandorfer et al. 2004) at the THEMIS solar telescope to observe Fe I 524.7 and 525.0 nm spectral lines at various heliocentric angles, reported that weak internetwork fields tend to adopt a horizontal orientation beyond $\mu = 0.2$. Here, $\mu$ represents the cosine of the heliocentric angle, and as $\mu$ decreases, progressively higher layers of the photosphere are sampled. Investigations utilizing data from the Solar Optical Telescope aboard the Hinode satellite (SOT; T. Kosugi et al. 2007; K. Ichimoto et al. 2008; T. Shimizu et al. 2008; Y. Suematsu et al. 2008; S. Tsuneta et al. 2008) around 630 nm have provided additional insight. T. A. Carroll & M. Kopf (2008) observed a fourfold decrease in the unsigned averaged flux of weak fields from the photosphere's bottom to a height of 500 km. B. Viticchié et al. (2011) asserted a swift vertical decline in the magnetic field, identifying kG fields at the photosphere's base and sub-kG strengths at higher layers. The spatially coupled inversions conducted by S. Danilovic et al. (2016) also confirmed a trend of magnetic fields weakening with increasing height. Complementary results from MHD simulations, such as those by M. Schüssler & A. Vögler (2008) and M. Rempel (2014), also support these findings. Both simulations indicate that the magnetic field in the internetwork is nearly isotropic in the deeper layers of the photosphere. As one moves to higher layers, the horizontal components of the magnetic field become increasingly dominant, with the ratio of horizontal to vertical field peaking at approximately 450 km above $\tau = 1$ (M. Rempel 2014). The clear dominance of the horizontal fields in the mid-photosphere seems to be a distinct characteristic of the strongly intermittent field produced by near-surface turbulent dynamo action, consistent with the assumption of a simple loop topology with a preferred length scale (M. Schüssler & A. Vögler 2008).

Another related topic that has garnered significant attention is the asymmetry in Stokes $V$ profiles across different solar regions. Stokes $V$ asymmetries can be observed on almost any part of the solar surface where a magnetic field is present, from sunspots to internetwork. The prevailing consensus, as posited by R. M. E. Illing et al. (1975), attributes these asymmetries to gradients in the magnetic field and velocity along the line of sight. Stokes $V$ asymmetries observed in network or plage regions have attracted the attention of solar physicists since their first detection in the 1980s (see, e.g., J. O. Stenflo et al. 1984). In the work of V. Martínez Pillet et al. (1997), a comprehensive examination of the characteristics of Stokes $V$ profiles in plage regions is provided. Subsequent efforts focused on unraveling the thermodynamic and magnetic properties of magnetic flux tubes, with numerous models proposed to elucidate the observed asymmetries in Stokes $V$ profiles within network patches (see U. Grossmann-Doerth et al. 1988; S. K. Solanki 1989; J. Sanchez Almeida et al. 1996; L. R. Bellot Rubio et al. 2000, and references therein). Among the models proposed, the thin flux-tube model assumes a predominantly vertical magnetic tube expanding with height (U. Grossmann-Doerth et al. 1988; S. K. Solanki 1989), while the MISMA hypothesis posits that the resolution element comprises an infinite number of magnetic atmospheres (J. Sanchez Almeida et al. 1996).

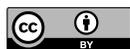

Advances in astrophysics often go hand in hand with improvements in both the instruments for performing observations and the techniques for diagnosing those observations. The evolution of solar observational capabilities has been pivotal in advancing our understanding of quiet Sun magnetism. The recent deployment of state-of-the-art solar telescopes, such as the Daniel K. Inouye Solar Telescope (DKIST; T. R. Rimmele et al. 2020), marks a significant milestone in this regard. Equipped with advanced instrumentation, DKIST offers unprecedented spatial resolution and spectral coverage, enabling detailed investigations into solar magnetic fields with unparalleled precision. One of the key methodologies employed in analyzing solar magnetism is multiline inversions of spectropolarimetric data. Multiline inversions involve the simultaneous interpretation of multiple spectral lines, allowing for a comprehensive characterization of magnetic fields, temperature, and velocity throughout the solar atmosphere. In recent years, significant progress has been made in understanding the magnetic properties of the quiet Sun using multiline inversions (C. Quintero Noda et al. 2017; T. L. Riethmüller & S. K. Solanki 2019; C. Kuckein et al. 2021; J. C. Trelles Arjona et al. 2021a). These studies have revealed intricate details about the spatial distribution of magnetic fields, the thermodynamic structure of the solar atmosphere, and the dynamics of plasma motions. Such findings have deepened our appreciation for the complex nature of solar magnetism and its influence on solar phenomena.

In this paper, we use both DKIST spectropolarimetric data and multiline inversions to perform a study of quiet Sun magnetism focused on its depth evolution. Our approach involves a meticulous computation of the log ($gf$) values corresponding to the spectral lines within the vicinity of the 630.1 nm spectral range, as observed by the ViSP instrument at DKIST solar telescope. Subsequently, we rigorously explore an inversion strategy that optimally incorporates all these spectral lines, specifically to untangle the intricate details of internetwork magnetism. Finally, we apply this refined inversion strategy to real observational data, aiming to address lingering questions and contribute valuable insights to this fundamental branch of solar magnetism.

## 2. Observations

This work used spectropolarimetric data obtained on 2022 October 24, between 19:47 and 20:39 UT, with the Visible Spectro-Polarimeter (ViSP; A. G. de Wijn et al. 2022) attached to the DKIST (T. R. Rimmele et al. 2020). The central helioprojective coordinates were approximately 198″ north and 398″ east, resulting in an angle of $\mu$ of 0.89. The scanning area was 75″.1 × 53″.0, along the slit and scan directions, with a step size and slit width of 0″.0534 and 0″.0536, respectively. We selected the part of the scan where the AO worked properly. Therefore, the dimensions of the data analyzed in this work are 75″.1 × 21″.3. The recorded spectral range was centered around 630.1 nm, spanning 1.25 nm (specifically from 629.55 to 630.80 nm in this data set). The spectral sampling and resolution were 12.83 mÅ and 22.00 mÅ, respectively. The total integration time per pixel and slit position was 480 ms to achieve a signal-to-noise ratio of 1100 in Stokes $I$ and 1000 in Stokes $Q$, $U$, and $V$. The adaptive optics system allowed for a spatial resolution of around 0″.15.

We used Level 1 data that were publicly released in 2023 April. However, further reduction processes were implemented: the subtraction of wavelength-independent stray-light contamination from the intensity spectra, the correction of residual crosstalk from Stokes $I$ into Stokes $Q$, $U$, and $V$, and the application of principal component analysis to reduce noise in the data (PCA M. M. Loève 1955; D. Rees & Y. Guo 2003; see J. C. Trelles Arjona et al. 2021b, and references therein for more details on the reduction process).

To show the high quality of the ViSP data, we display in Figure 1 the intensity and polarization maps of the observed data set after the additional reduction process. Although the left panels show the internetwork, many polarimetric signals (both in linear and circular polarization) can be seen. The internetwork panels (right column in Figure 1) were rescaled to avoid saturation for the sake of visualization.

### 2.1. Determination of log (gf)

Under specific conditions, multiline inversions can yield more accurate results in determining the thermodynamic and magnetic properties of the solar atmosphere (T. L. Riethmüller & S. K. Solanki 2019), albeit demanding greater computational efforts. An essential aspect of multiline inversions involves accurately characterizing all spectral lines involved. Among the crucial parameters in spectral line characterization is the oscillator strength, which is defined as the ratio of the probability of electromagnetic radiation during a transition between atomic or molecular energy levels to the probability of the corresponding transition in a classical oscillator (D. Mihalas 1978). In astrophysics, the oscillator strength is conventionally treated through the log ($gf$) parameter, where $f$ represents the oscillator strength itself and $g$ denotes the statistical weight of that transition.

The spectral range observed by the ViSP instrument around 630.1 nm contains the pair of spectral lines that have long been used and well tested by the solar community at 630.15 and 630.25 nm. However, within this spectral range, other spectral lines are less rigorously examined. To assess the accuracy of log ($gf$) values for these less-tested spectral lines, we employed the Stokes Inversion based on the Response functions inversion code (SIR, B. Ruiz Cobo & J. C. del Toro Iniesta 1992). SIR solves the radiative transfer equation for polarized light assuming local thermodynamic equilibrium and the Zeeman effect. These approximations are very suitable for most photospheric lines, such as those around 630.1 nm. SIR seeks synthetic profiles that better match the observed ones. To do so, SIR minimizes the quadratic distance between the synthetic and observed profiles by means of the Levenberg–Marquard iteration scheme.

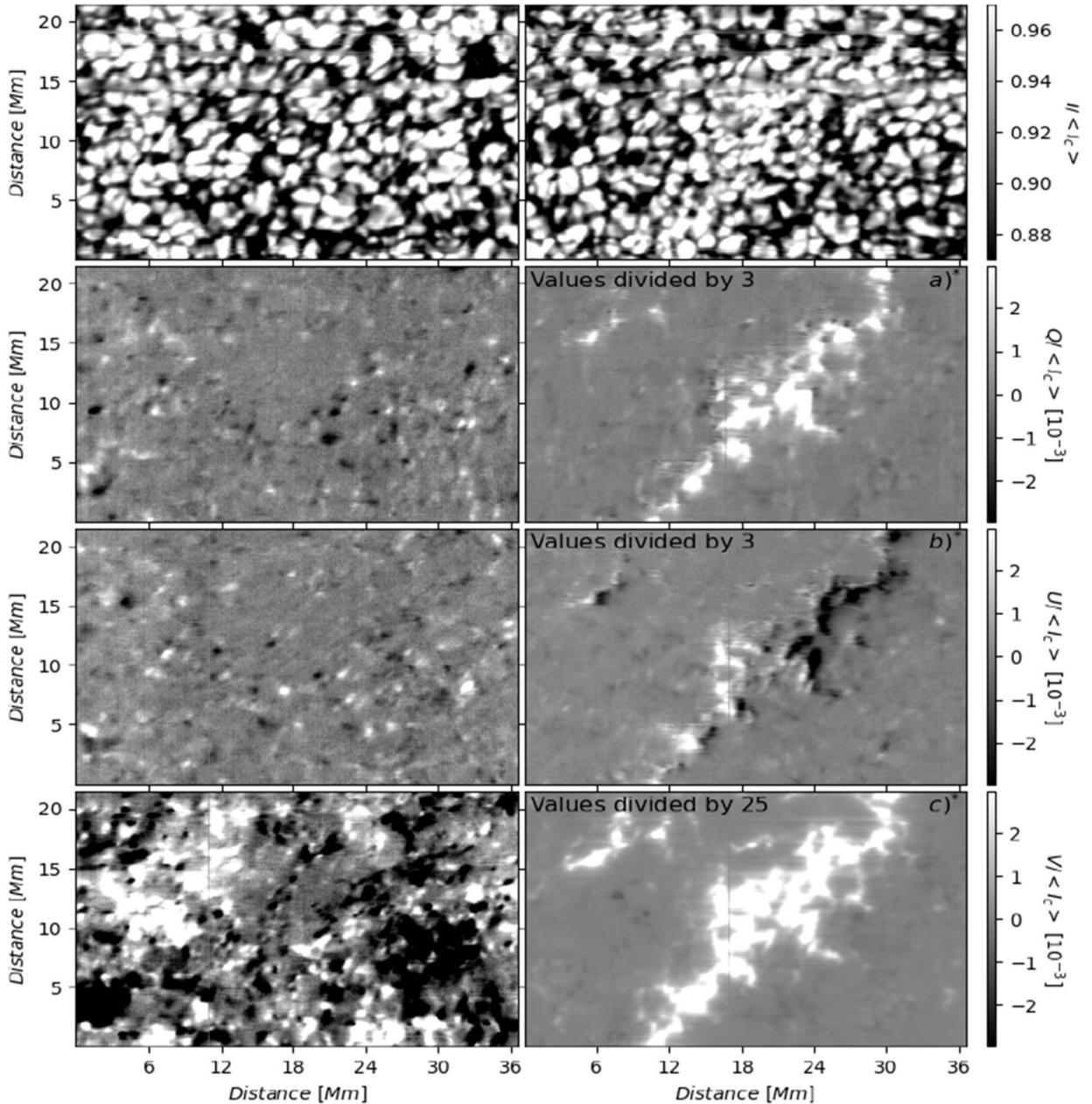

**Figure 1.** Observations used in this work. From top to bottom, we display the continuum of the intensity and the amplitudes of Stokes $Q$, $U$, and $V$ in a single wavelength point (630.321 and 630.244 nm for continuum and polarimetry panels, respectively). Panels (a), (b), and (c), marked with an asterisk, are changed for visualization purposes to see the network patch. The signal in panels (a) and (b) is divided by 3. The signal in panel (c) is divided by 25.

Despite initially using log ($gf$) values from literature sources (R. L. Kurucz & B. Bell 1995) for inversions, they did not yield satisfactory fits to observed spectral lines. Attempts to simultaneously invert all spectral lines using standard atmospheric models resulted in inadequate fits. Further refinement of fits in one spectral line led to degraded fits in the remaining spectral lines. Consequently, we opted to determine log ($gf$) values independently to enhance their accuracy.

To determine the log($gf$) values of all the spectral lines we used the method reported in J. C. Trelles Arjona et al. (2021b). This method is an iterative process to find the best combination of log($gf$) values and model atmospheres to minimize the difference between the synthesized and observed spectral profiles. We took abundances from M. Asplund et al. (2009) for the whole process.

Initially, we used the average quiet Sun spectra from the ViSP observations described above to determine the log($gf$) values of the three strongest spectral lines of the spectral range: Fe I at 629.78, 630.15, and 630.25 nm. In the inversion,

**Table 1**
Atomic Line Parameters of the 630.1 nm ViSP Spectral Band

| Element | Wavelength (nm) | $J_l$ | $J_u$ | $\log(gf)$ | $\log(gf)^*$ | $\chi_e$ (eV) | $g_{\text{eff}}$ | $\alpha$ | $\sigma$ ($a_0^2$) |
|---|---|---|---|---|---|---|---|---|---|
| Fe I | 629.78 | 1.0 | 2.0 | $-2.65 \pm 0.01$ | $-2.74$ | 2.221 | 1.00 | 0.264 | 278 |
| Sc II | 630.07 | 2.0 | 2.0 | $-1.99 \pm 0.02$ | $-1.84$ | 1.507 | 1.33 | ⋯ | ⋯ |
| Fe I | 630.15 | 2.0 | 2.0 | $-0.56 \pm 0.01$ | $-0.67$ | 3.640 | 1.67 | 0.243 | 837 |
| Fe I | 630.25 | 1.0 | 0.0 | $-1.12 \pm 0.01$ | $-1.13$ | 3.690 | 2.50 | 0.241 | 1062 |
| Fe I | 630.35 | 6.0 | 5.0 | $-2.46 \pm 0.03$ | $-2.66$ | 4.317 | 1.51 | 0.279 | 731 |
| Ti I | 630.38 | 3.0 | 3.0 | $-1.28 \pm 0.02$ | $-1.57$ | 1.442 | 0.92 | 0.236 | 353 |

**Note.** The columns show (from left to right) the element and the ionization state, wavelength, total angular momentum quantum number of the lower and upper levels ($J_l$ and $J_u$), the oscillator strength $\log(gf)$ determined in this work, the oscillator strength $\log(gf)^*$ extracted from R. L. Kurucz & B. Bell (1995), the excitation potential of the lower level ($\chi_e$), the Landé g-factor, and the coefficients for collisional broadening ($\alpha$ and $\sigma$). Abundances: M. Asplund et al. (2009) (7.50, 4.95, and 3.15 dex for Fe, Ti, and Sc respectively).

we used as the initial atmosphere the atmosphere of the FAL-C model (J. M. Fontenla et al. 1993). Throughout the entire $\log(gf)$ determination process, we worked with the spectral sampling of the ViSP data used in the study, specifically 12.83 mÅ. To account for ViSP spectral resolution, a convolution with a Gaussian profile of $\sigma = 22.0$ mÅ was applied to synthesize the spectral profiles. This value was inferred directly from the ViSP data used in this study. Subsequently, we used the previously determined $\log(gf)$ values to perform pixel-by-pixel inversions of the observed region. Thus, we obtained model atmospheres that can be used to determine the $\log(gf)$ values of the remaining spectral lines. Then, we performed inversions pixel by pixel using the derived $\log(gf)$ values of all spectral lines. Finally, we again determined $\log(gf)$ values and model atmospheres pixel by pixel. The process allowed us to determine the mean and standard deviation of $\log(gf)$ for all spectral lines, averaging the results over a large number of pixels to minimize biases from individual atmospheric conditions and inversion procedures. It is worth noting that there are additional sources of uncertainty beyond the standard deviation, including spectral noise, systematic biases in the inversion process, and degeneracies between the $\log(gf)$ values and the atmospheric profile. While these factors may contribute to the overall uncertainty, we believe their impact has been mitigated by the large number of determinations performed for each value. A comprehensive evaluation of these effects would benefit from controlled experiments using synthetic data, where the true atmospheric parameters and line properties are known. While such a study is beyond the scope of the present work, it would represent a valuable next step to further validate and quantify the robustness of the $\log(gf)$ determinations.

The final results are shown in Table 1. It was not possible to calculate the collisional broadening coefficients $\alpha$ and $\sigma$ for the Sc II line at 630.07 nm. In Table 1, they appear as values equal to 0.0. In these cases, SIR uses the classical Unsöld equation (A. Unsold 1955) to calculate the damping factor.

### 3. Testing Stokes Multiline Inversions with Numerical Simulations

To search for the best inversion strategy to invert ViSP data around 630.1 nm, we have synthesized and inverted the spectral lines listed in Table 1 using the Stokes inversion based on the response functions inversion code (SIR; B. Ruiz Cobo & J. C. del Toro Iniesta 1992) in a MANCHA3D magnetohydrodynamical simulation (E. Khomenko et al. 2017).

We used the quiet Sun simulation performed by the MANCHA3D code (E. Khomenko & M. Collados 2006; T. Felipe et al. 2010). MANCHA3D code solves the equation of nonideal magnetohydrodynamics, along with a realistic equation of state and nongray radiative transfer (E. Khomenko & M. Collados 2012). In the simulation, the dynamo action amplifies an initial seed magnetic field (provided from the Bierman battery term). It is important to us that the magnetic field values that can be found in the simulation are on the order of those inferred from solar observations (see E. Khomenko et al. 2017 for details of the numerical setup of the simulation).

Traditionally, in this spectral range, inversions are performed only with the two spectral lines at 630.15 and 630.25 nm, which are highly sensitive to magnetic fields (L. R. Bellot Rubio & D. Orozco Suárez 2019). This pair of spectral lines has been widely used for many years in solar physics to perform inversions. They are the lines observed by SOT (T. Shimizu et al. 2008) on board the HINODE satellite (T. Kosugi et al. 2007), which has provided excellent results to the solar community. The spectral range around 630.1 nm observed by the ViSP instrument is 1.25 nm wide. Thus, we have access to more spectral lines that can be used in inversions, with the benefits that this entails, increasing, for instance, the spatial resolution in depth (C. Quintero Noda et al. 2017). Therefore, we have performed a series of experiments to check the benefits of multiline inversions and to find the inversion strategy that provides the best results in such a spectral range.

The SIR code uses a node-based approach to determine the stratification of the atmosphere. They are locations along the optical depth where SIR allows perturbations to the atmospheric parameters. SIR interpolates in between nodes either with low-order polynomials or with splines. Hence, the greater the number of nodes, the more complexity in the gradients. Multiline inversions let us increase the number of nodes. Increasing the number of nodes expands the degrees of freedom in the solution. In contrast, introducing additional constraints (such as those imposed by the inclusion of multiple spectral lines) effectively reduces the degrees of freedom. Consequently, increasing the available information (e.g., by incorporating additional spectral lines) enables the use of a greater number of nodes without a proportional increase in the degrees of freedom. Thus, for temperature, line-of-sight (LoS) velocity, and magnetic field strength and inclination, we let the SIR code automatically choose in each pixel the number of nodes by evaluating the amount of information contained in the Stokes profiles (see more details in J. C. del Toro Iniesta &

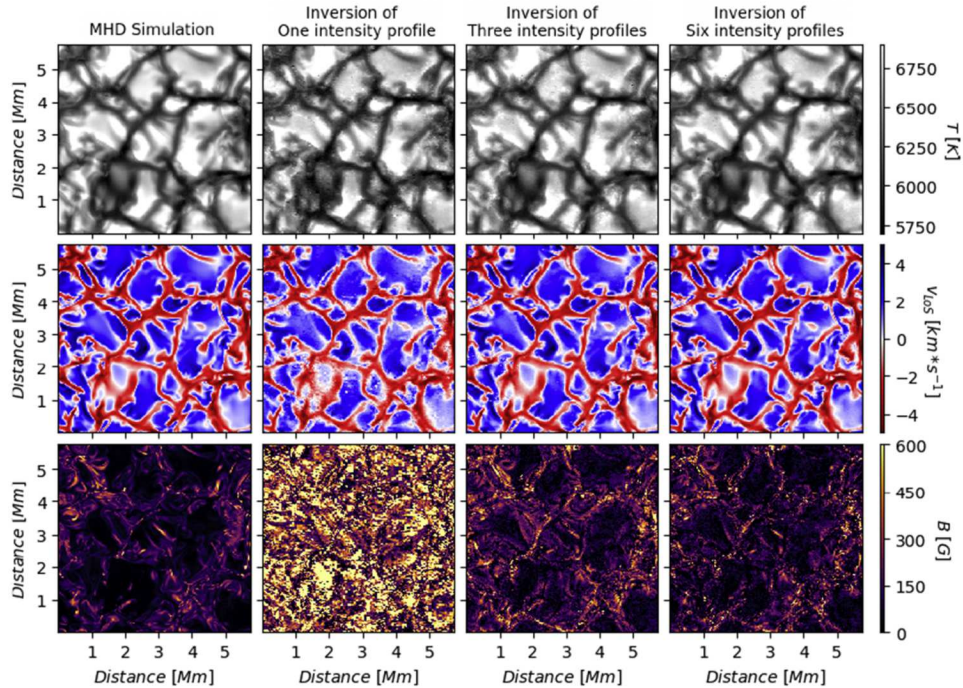

**Figure 2.** Comparison between the simulation (first column from the left) and the results achieved by the inversion of one (Fe I at 630.25 nm), three (Fe I lines at 630.15, 630.25, and 630.35 nm), and six (all lines in Table 1) spectral lines (second, third, and fourth columns from the left, respectively). Temperature (at log $(\tau_{500}) = 0.0$), LoS velocity (at log$(\tau_{500}) = -0.4$) and magnetic field strength averaged over layers with the maximum Pearson correlation between the simulation and the inversion (from log$(\tau_{500}) = -0.2$ to log$(\tau_{500}) = -0.6$) in top, middle, and bottom rows, respectively.

B. Ruiz Cobo 2016). On the other hand, we consider that the information is still not enough to set the free selection of nodes in the case of magnetic field azimuth. Therefore, we select two nodes for this parameter to have an idea of the gradients where polarimetric signals are enough to capture their behavior.

Even though the complexity of gradients in the simulation is high, the SIR code is capable to infer the global behavior of magnetic field vector gradients. Therefore, in all the upcoming plots we have worked with averaged magnetic field vector quantities (strength, inclination, and azimuth). The averaged range in depth, from log$(\tau_{500}) = -0.2$ to log$(\tau_{500}) = -0.6$, has been deduced from the response functions and the Pearson correlation between the simulation and the results of the inversions. The symbol $\tau_{500}$ denotes the optical depth at 500 nm. For the temperature and LoS velocity plots we selected just one optical depth (log$(\tau_{500}) = 0.0$ for temperature and log$(\tau_{500}) = -0.4$ for the LoS velocity). These optical depths were deduced as the maximum Pearson correlation between the simulation and the results of the inversions.

### 3.1. Full Resolution, Noiseless Case

In J. C. Trelles Arjona et al. (2021a), we demonstrated that, under specific conditions, using only intensity profiles in the inversions is not only viable, but also advisable. In this work, our initial investigation revolved around assessing the feasibility of measuring magnetic field strength solely through the inversion of intensity profiles within the spectral range around 630.1 nm. Furthermore, we conducted an experiment to examine the impact of the number of spectral lines included in the inversion on the results. For this purpose, we employed original, noise-free simulations with the original spatial resolution. In this particular experiment, we omitted the inversion of the microturbulence velocity, presuming that the velocities along the line of sight are resolved.

The outcomes of the conducted experiment are illustrated in Figure 2. From this point onward, one-line inversions refer to those performed with the 630.25 nm line, three-line inversions correspond to those using the 630.15, 630.25, and 630.35 nm lines, and six-line inversions include all the lines listed in Table 1. A rapid visual assessment reveals that, particularly in the estimation of magnetic field strength, an increase in the number of spectral lines used in inversions leads to more accurate results.

In Figure 13 (shown in the Appendix) we display the results in depth. Each star represents the Pearson correlation between the simulation and the inversion for each optical depth. The plot underscores the trend for improved results with an increase in the number of spectral lines in the inversion. Notably, the enhancement from using three lines to six lines is already subtle. Consequently, the inference of the magnetic field strength encounters limitations imposed by the spectral range. For comparison purposes, the results obtained at 1.5 $\mu$m with 15 spectral lines are depicted by red stars (J. C. Trelles Arjona et al. 2021a). At 1.56 $\mu$m, the averaged Pearson's $r$ correlation coefficients for magnetic field strength, temperature, and LoS velocity are 0.635, 0.964, and 0.978, respectively.

As expected in this experiment, results at longer wavelengths exhibit greater accuracy due, among other factors, to the quadratic dependence of Zeeman splitting on wavelength. The comparison between these two spectral ranges confirms that infrared lines provide higher accuracy in the inferred parameters. This advantage is particularly evident in deeper layers, as spectral lines around 1565 nm are formed lower in the solar atmosphere compared to those at 630.1 nm. The analysis conducted by J. M. Borrero et al. (2016) supports this, showing that the continuum at 1565 nm forms 30–70 km deeper than at 630 nm, depending on the solar region. Furthermore, response function calculations demonstrate that the 1565 nm lines are significantly more sensitive to the magnetic field at $\log(\tau) = 0.0$ and provide a more localized response, whereas the 630 nm lines distribute their sensitivity over a broader range of optical depths (J. M. Borrero et al. 2016). Another key factor is the greater variation in excitation potential, $\log(gf)$, and the Landé factor among the infrared lines compared to the visible ones. For example, in the infrared range, Fe I 1564.85 nm has a Landé factor of 3, while the strong Fe I 1566.53 nm line has a Landé factor of 0.725. In contrast, in the visible range, the Landé factor varies from 2.5 for Fe I 630.25 nm to 0.92 for Ti I 630.38 nm (J. C. Trelles Arjona et al. 2021b).

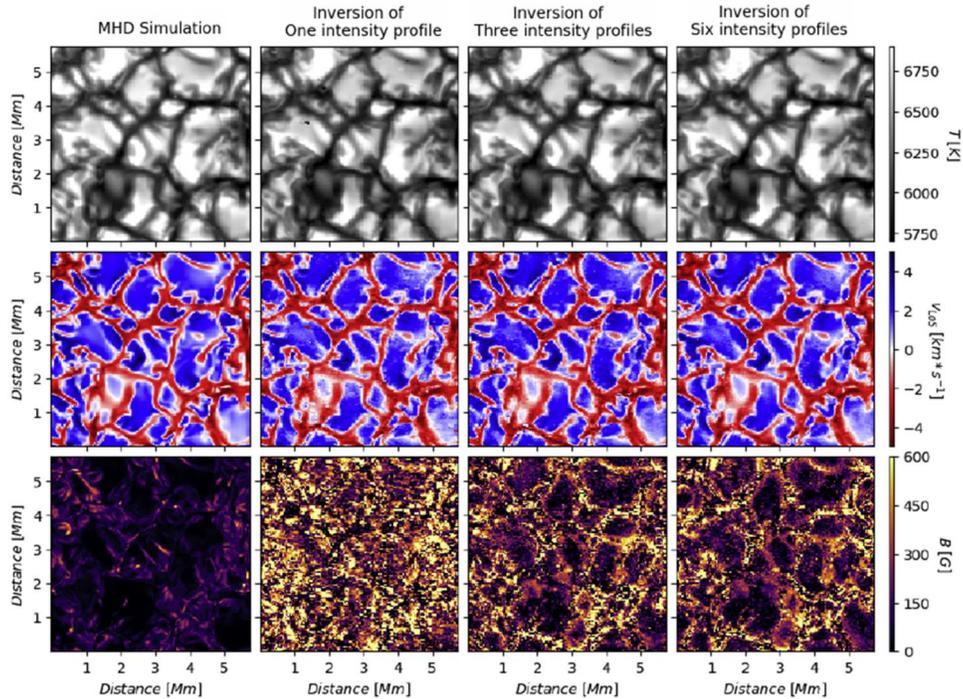

**Figure 3.** Comparison between some parameters of the simulation (left column) averaged within the larger pixel to match the pixel size of ViSP observations, and the same parameters inferred from the inversion of one (Fe I at 630.25 nm), three (Fe I lines at 630.15, 630.25, and 630.35 nm), and six (all lines in Table 1) Stokes $I$ profiles (second, third, and fourth columns from the left, respectively). From top to bottom: temperature (at $\log(\tau_{500}) = 0.0$), LoS velocity (at $\log(\tau_{500}) = -0.4$), and magnetic field strength averaged along the layers with maximum Pearson correlation between the simulation and the inversion (from $\log(\tau_{500}) = -0.2$ to $\log(\tau_{500}) = -0.6$).

### 3.2. Spatially Degraded, Noisy Case

The emergence of the new generation of 4 m aperture telescopes presents an opportunity for increased spatial resolution of the solar surface observations. However, achieving spatial resolution comparable to the original simulation remains a challenging goal. Until this level of resolution is attained, the challenge lies in recovering the characteristics of the magnetic field vector when the fields are not fully resolved. To replicate the loss of information resulting from the finite spatial resolution in current state-of-the-art spectropolarimetric data, the synthetic profiles obtained from the simulations are rebinned to match the ViSP pixel size (approximately 40 km) (ViSP, A. G. de Wijn et al. 2022). It is important to note and bear in mind that rebinning simulation atmospheres is not equivalent to rebinning synthetic profiles derived from those simulations. A mixture of velocities is produced when rebinning. Hence, from now on in the inversions, we allow for variations with depth of the microturbulence velocity parameter in subsequent inversions. In addition, Gaussian noise is added and the data are spectrally convolved with a Gaussian profile of $\sigma = 22.0$ mÅ (as inferred from the ViSP data set used in this study) to simulate realistic observational conditions of the ViSP instrument.

To find out the optimal inversion strategy suited for the data acquired by the ViSP instrument at 630.1 nm, we conducted three experiments.

#### 3.2.1. Experiment 1: Inversion of Intensity Profiles with Different Number of Spectral Lines

In the initial experiment, considering the outcomes in 3.1, we replicated the inversion using only the intensity profiles, incorporating the data treatment discussed earlier. The results are presented in Figures 3 and 14. A comparison between Figures 2 and 3 reveals a substantial impact of the combined effects of resampling and noise addition, which considerably degrade the accuracy of magnetic field strength inference. This effect is further evident in Figure 14 (shown in the Appendix), where the results for temperature and LoS velocity exhibit a slight deterioration, while the degradation in magnetic field strength is more pronounced. Furthermore, the inclusion of more spectral lines in the inversion continues to demonstrate an improvement in the results.

#### 3.2.2. Experiment 2: Inversion of Full Stokes Profiles with Different Number of Spectral Lines

The outcomes in Section 3.2.1 led us to incorporate polarimetry in the subsequent experiment to enhance the accuracy of magnetic field strength measurements. In this experiment, we replicated Experiment 1 but included the Stokes $Q$, $U$, and $V$ profiles. A comparison between Figures 3 and 4 highlights the notable improvement in the inference of magnetic field strength. However, there is an unexpected deterioration in temperature and LoS velocity results. This contrast behavior is evident in the comparison between Figures 14 and 15 (shown in the Appendix). It is worth noting that the experiments conducted with the simulations were all performed using the same inversion setup. Specifically, the number of initializations was set to seven in all cases. The results presented in this section demonstrate that, given the setup of these experiments (i.e., typical internetwork magnetic fields observed with the ViSP instrument), introducing the Stokes parameters $Q$, $U$, and $V$ increases the likelihood of converging to a local minimum rather than the desired global minimum.

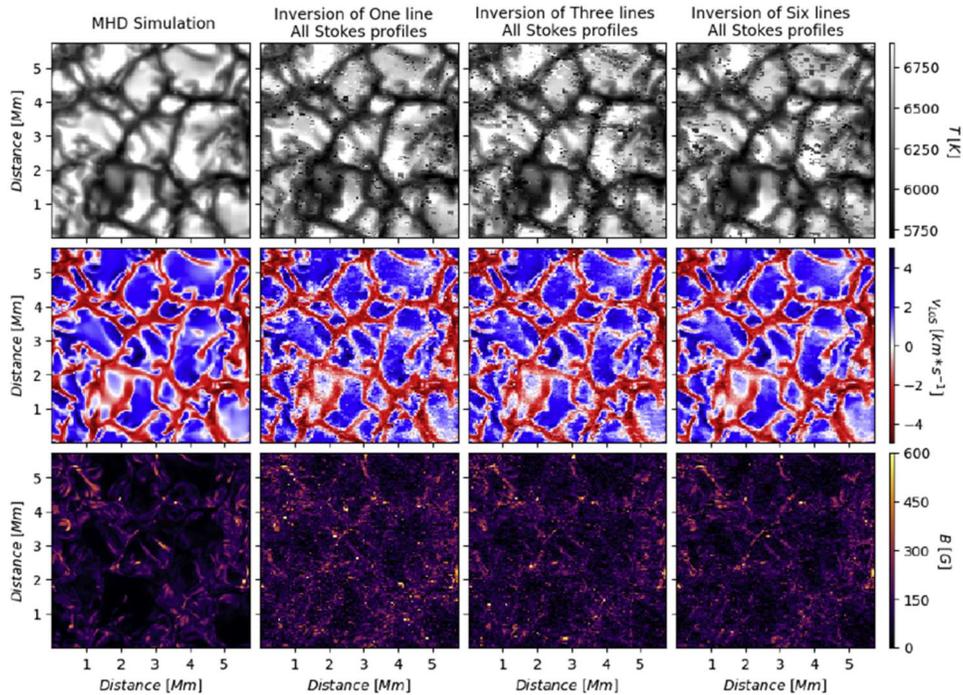

**Figure 4.** Comparison between some parameters of the simulation (left column) averaged within the larger pixel to match the pixel size of ViSP observations, and the same parameters inferred from the inversion one (Fe I at 630.25 nm), three (Fe I lines at 630.15, 630.25, and 630.35 nm), and six (all lines in Table 1) full Stokes profiles (second, third, and fourth columns from the left, respectively). From top to bottom: temperature (at $\log(\tau_{500}) = 0.0$), LoS velocity (at $\log(\tau_{500}) = -0.4$), and magnetic field strength averaged along the layers with maximum Pearson correlation between the simulation and the inversion (from $\log(\tau_{500}) = -0.2$ to $\log(\tau_{500}) = -0.6$).

*3.2.3. Experiment 3: Inversion of Six Intensity Profiles and Different Number of Spectral Lines in Stokes Q, U, and V*

In Sections 3.2.1 and 3.2.2, we have demonstrated that the optimal strategy to infer the magnetic field strength involves using six spectral lines in all Stokes parameters. However, this strategy adversely affects results in temperature and LoS velocity. Thus, in this experiment, we used the outcomes in Section 3.2.1 as initial atmospheres for a second inversion, where neither temperature nor LoS velocity were inverted. Throughout this experiment, we used six intensity profiles, altering the number of spectral lines in the Stokes $Q$, $U$, and $V$ profiles (specifically, one, three, and six spectral lines).

Figures 5 and 16 show the results of this experiment. A cursory examination of Figure 5 reveals that the strategy with all lines yields noisier results, whereas the results with one and three lines exhibit improvement. An additional noteworthy aspect in Figure 5 is that, under the specific conditions of the experiments conducted in this study (i.e., inversion strategies, spectral and spatial resolution of simulated data, noise levels, etc.), we are still unable to accurately determine the azimuth of the magnetic field. This limitation is likely influenced by the additive noise introduced in the simulation, and improved results may be achievable with higher signal-to-noise observations.

This limitation is corroborated by the Pearson correlation depicted in Figure 16 (shown in the Appendix), where the best results are achieved with the one-line and three-line strategies. Results with one and three lines exhibit a similar performance. To properly compute the Pearson correlation coefficient between the inferred values and the azimuth values that come from the simulation, we accounted for the 180° ambiguity. For each pixel, we evaluated both the inferred azimuth and its 180° equivalent, selecting the value that minimized the difference with the simulation.

In Figure 6, we present the final results obtained through the six-intensity and three-polarized profiles inversion strategy for selected atmospheric parameters as a function of optical depth. The plots depicting magnetic field strength ($B$) and inclination ($\gamma$) on the lower left and lower right, respectively, showcase our ability to capture the behavior with depth in both cases. The recovery of magnetic field strength behavior with depth is well determined where the sensitivity of the spectral lines of this spectral domain is high ($\log(\tau_{500}) = -0.2$ to $\log(\tau_{500}) = -1.2$). Notably, in the magnetic field inclination plot, we have segregated pixels on the basis of polarity before averaging. The inversion results, represented by the green solid line, exhibit a tendency to depict magnetic fields as more horizontal than those presented in the simulation (blue solid line), primarily influenced by photon noise effects (J. M. Borrero & P. Kobel 2012). Despite this, we are able to determine the general trend in depth. This instills confidence in the application of this inversion strategy to real data, allowing us to uncover the general behavior of the magnetic field with depth.

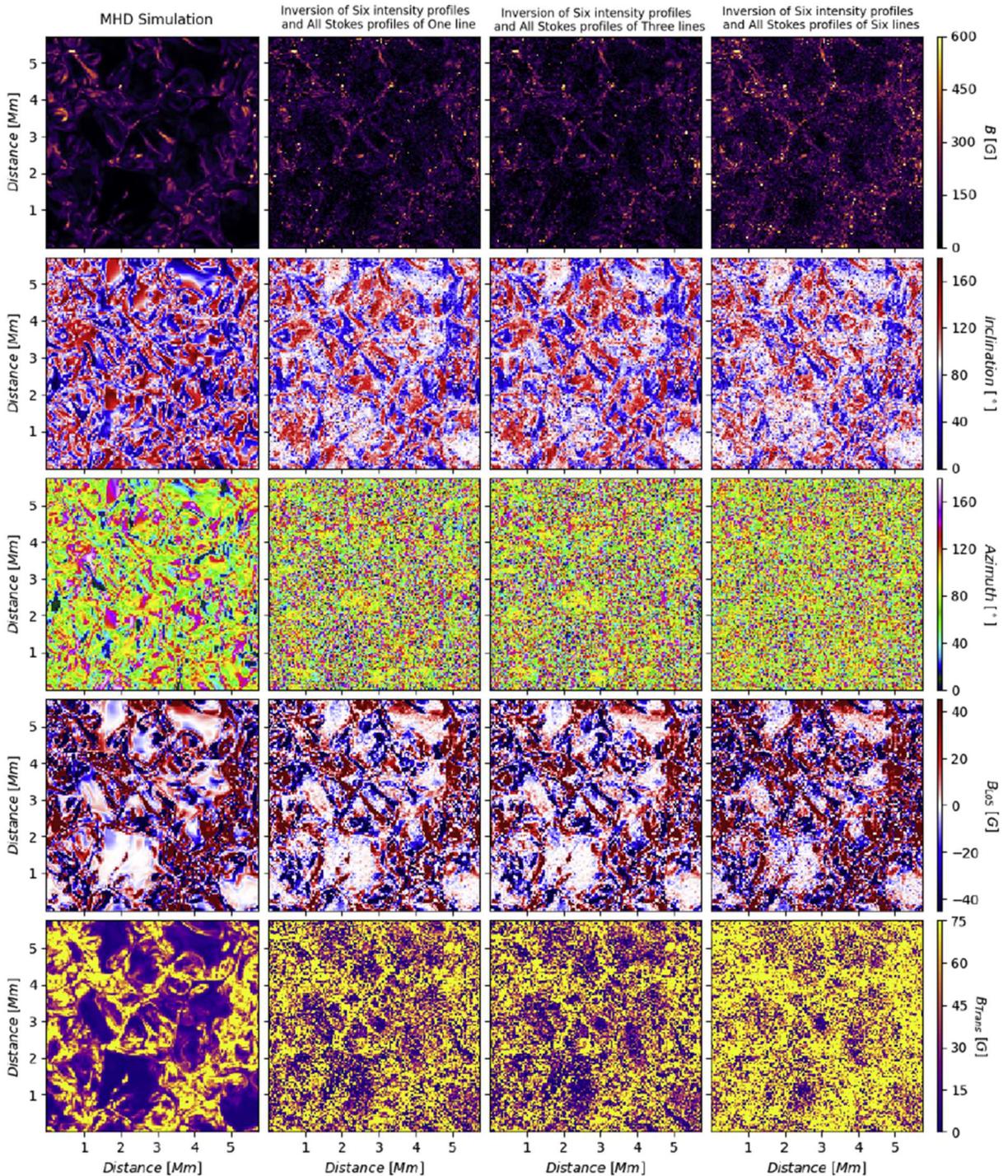

**Figure 5.** Comparison between some parameters of the simulation (left column) averaged within the larger pixel to match the pixel size of ViSP observations, and the same parameters inferred from the inversion of Stokes $I$, $Q$, $U$, and $V$ profiles simultaneously. In the inversions, six spectral lines are used in intensity profiles and one (Fe I at 630.25 nm), three (Fe I lines at 630.15, 630.25, and 630.35 nm), and six (all lines in Table 1) spectral lines in Stokes $Q$, $U$, and $V$ (second, third, and fourth columns from the left, respectively). From top to bottom: magnetic field strength, inclination, azimuth, and LoS and transverse magnetic flux. All quantities averaged along the layers with maximum Pearson correlation between the simulation and the inversion (from $\log(\tau_{500}) = -0.2$ to $\log(\tau_{500}) = -0.6$).

It should be noted that, in general, employing the strategy of utilizing all six spectral lines in Stokes $Q$, $U$, and $V$ yields better results in pixels characterized by higher magnetic field strength values. Conversely, more favorable outcomes are obtained by incorporating fewer Stokes $Q$, $U$, and $V$ profiles in pixels exhibiting lower magnetic field strength values.

The implications derived from these experiments appear clear and well supported. The integration of additional spectral lines in inversions consistently improves the results, particularly improving spatial resolution in depth. In the context of our investigation, focused on internetwork magnetism, optimal outcomes are achieved by including all intensity profiles and only the most intense spectral lines in polarimetry (specifically, three in this spectral range). This approach is justified by the adverse effects we found when introducing Stokes profiles without discernible signals (i.e., containing only noise), which lead us to achieve noisy results. It is worth mentioning that, to select the best fit among all initializations for each pixel, we use the reduced chi-squared statistic as our merit function. In this approach, we assign a weight to each Stokes parameter based on its signal-to-noise ratio. A potentially improved strategy would be to assign individual weights to each spectral line within each Stokes profile. This would allow the inclusion of all spectral lines across the four Stokes parameters while down-weighting or entirely excluding those profiles dominated by noise. This alternative approach could lead to even more accurate results. However, it is important to consider the computational cost of including additional spectral lines in the inversion process when, for a significant fraction of the pixels, they would contribute little or no useful information. Increasing the number of initializations is another way to improve the results, as it increases the likelihood of reaching the global minimum. However, this also increases computational costs.

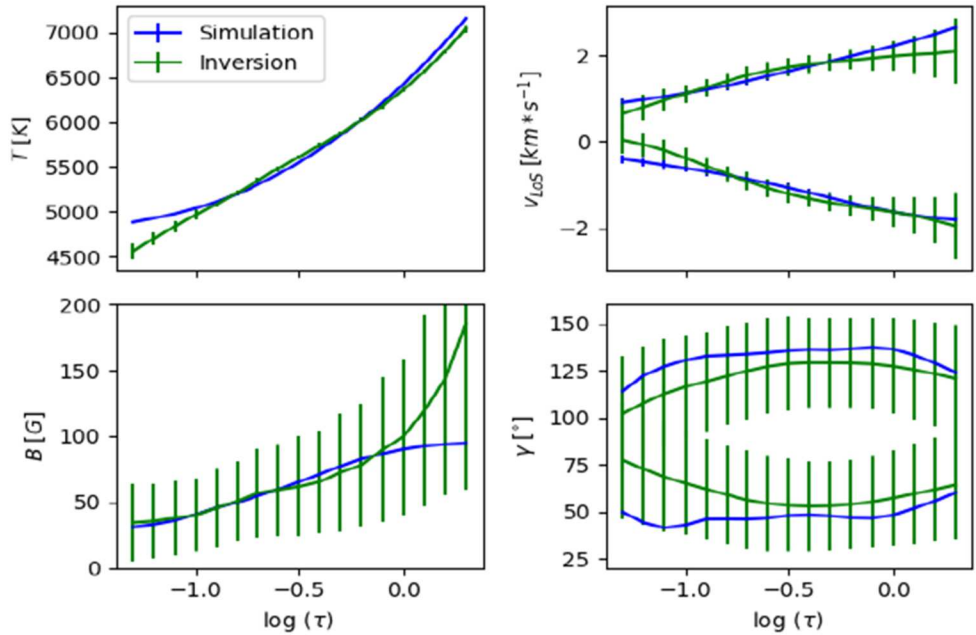

**Figure 6.** Results of the best inversion strategy applied to simulations as a function of the logarithm of the optical depth at a reference wavelength of 500 nm, log ($\tau_{500}$), to temperature $T$ (top left), LoS velocity $v_{LoS}$ (top right), the magnetic field strength $B$ (bottom left), and magnetic field inclination with respect to the observer line of sight $\gamma$ (bottom right). Blue solid lines correspond to the average of the simulation, while the green lines correspond to the average of the results of the inversion. The error bars represent the $\sigma$ value of the differences between the simulation and the inversion in each optical depth.

## 4. Inference of Physical Parameters From Real ViSP Data

After the simulation tests, we proceeded to perform inversions of the real observations described in Section 2. Based on the insights gained from the simulated experiments, our observational analysis incorporated six spectral lines in intensity and three in polarimetry (specifically, the most intense ones in this spectral range). The inversion procedure was executed through two sequential runs. In the initial run, only the intensity profiles were inverted, with free parameters including temperature, LoS velocity ($v_{LoS}$), microturbulence velocity, and magnetic field strength and inclination. Subsequently, in the second run, polarimetry was incorporated into the inversion process. In this phase, we retained the temperature and $v_{LoS}$ values obtained in the first run. Thus, the free parameters for the second run included magnetic field strength, inclination, azimuth, and microturbulence velocity. As in the simulation tests, we allowed the SIR code to autonomously determine the number of nodes for each pixel by evaluating the information contained in the Stokes profiles for all parameters except the azimuth of the magnetic field, where we selected two nodes. The initial atmosphere for the inversion was set as model C from J. M. Fontenla et al. (1993, FAL-C model). In the first run inversions, we introduced random variations to the initial model atmosphere in LoS and microturbulent velocities (within the range of 0–5 km s$^{-1}$) and magnetic field strength (within the range of 0–1500 G). For the second run, random modifications were applied to the initial model atmosphere in microturbulent velocity (within the range of 0–5 km s$^{-1}$) and in magnetic field strength and inclination (within the ranges of 0–1500 G and 0°–180°, respectively). A total of 11 inversions per pixel and per run were performed, and the solution with the lowest $\chi^2$ was selected. The adopted inversion strategy assumed the presence of one magnetic atmosphere occupying a fraction of the resolution element, with Stokes $I$ stray-light filling the remaining space.

### 4.1. Dealing with Stray Light

To achieve precise results in the inversion of spectro-polarimetric data, it is crucial to consider the impact of stray light. Stray light refers to the scattering of light from various regions of the source, causing photons from outside the Airy disk to fall onto a designated pixel. This introduces a fraction of photons that do not truly belong to that pixel but originate from other parts of the source.

Handling stray light involves addressing its shape and the quantity falling on each pixel. Recent approaches for calculating stray-light profiles have diverged, with one suggestion in D. Orozco Suárez et al. (2007) advocating local stray light (averaging around the target pixel), and another in A. Asensio Ramos & R. Manso Sainz (2011) proposing global stray light (considering all pixels in the observation). To decide between the two options, we performed inversions of real ViSP observations using the local stray-light approach, averaging around the target pixel in 0″.5, 1″, and 2″ boxes, along with the global stray-light approximation. The results indicated that the fits deteriorated starting from the 2″ box. Consequently, opting for a compromise between the 0″.5 and 1″ boxes, we chose the latter to mitigate the effects of selecting an excessively small local stray light (A. Asensio Ramos & R. Manso Sainz 2011).

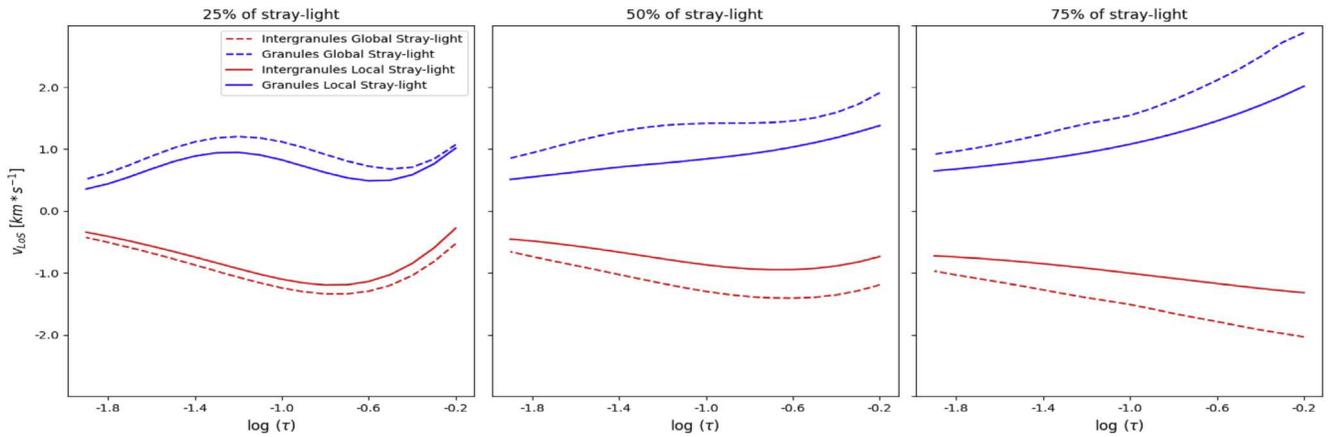

**Figure 7.** Dependence of velocity field on the amount of stray light (25%, 50%, and 75% from left to right). Blue solid and red dotted lines indicate the results of the inversions using local and global profile of stray light, respectively.

To determine the stray-light amount, we relied on the theory that the separated and averaged granular and intergranular velocities should accelerate with depth in the solar photosphere. However, the convective blueshift in the averaged profile obscured the behavior of intergranular velocities, particularly in deep layers. A reasonable approximation was found by increasing the stray-light percentage until the averaged intergranular velocities increased (in absolute value) with depth. Figure 7 displays variations in averaged velocities for granules (blue) and intergranules (red) in different percentages of stray light (25%, 50%, and 75%). A 1″ box around the pixel was used for both local (solid line) and global (dashed line) stray-light inversions to verify their influence on the stray-light amount calculation. The results indicated that a percentage of stray light of at least 75% was needed to maintain the expected acceleration with depth in intergranular velocities. The shape of the stray light did not influence the required amount, leading us to adopt the inversions using a stray-light profile averaged over a 1″ box around the pixel with a percentage of stray light of 75%. It is important to highlight the interplay between spatial resolution and stray light in the data. Although the calculated spatial resolution is 0″.15, this value corresponds to the best conditions in the data set (i.e., those slit positions where the adaptive optics system performed optimally). However, this resolution can vary significantly across the scan due to factors such as changing seeing conditions, variations in adaptive optics performance, defocus, and other effects. These variations likely explain the need to use a 1″ box around the pixel and the high stray-light fraction.

## 5. Results

Figure 8 illustrates the fit achieved to the observed profiles randomly selected in the internetwork (left) and the network (right). The quality of the fit is evident, although the spectral line at 629.78 nm presents challenges in linear polarization due to its low Landé factor compared to the lines at 630.15 and 630.25 nm. However, in network pixels characterized by a higher magnetic field strength, the fit remains satisfactory even for this spectral line.

In Figure 9, we present the surface result of the inversion at $\log(\tau_{500}) = -0.8$ for various atmospheric parameters. From top to bottom: magnetic field strength, inclination, azimuth, longitudinal and transverse magnetic field, and microturbulence velocity. A visual inspection reveals a discernible structure reminiscent of the granulation pattern across several atmospheric parameters. Separating the granular and intergranular zones based on temperature, we observe that in the intergranular zone, measurements show higher longitudinal and transverse magnetic field values (2.75 and 1.60 times, respectively) and microturbulence velocity (1.12 times) compared to the granular zone. The network's magnetic field exhibits a predominantly kiloGauss (kG) strength, with certain elements well above 2000 kG. The patch consistently maintains a uniform polarity, initially aligning almost entirely vertically at the center and gradually transitioning to a more horizontal orientation toward the radial peripheries. The magnetic field azimuth appears to divide the patch into two distinct regions, characterized by an angular separation of approximately 120°.

### 5.1. Evolution of Internetwork Magnetism with Depth

As demonstrated in Section 3, multiline inversions within the spectral range around 630.1 nm, as observed by ViSP at DKIST, provide valuable insights into the general behavior of internetwork magnetism with depth. The average stratification of magnetic field strength in the internetwork is presented in Figure 10 (bottom left). The magnetic field strength shows an evident decrease with decreasing optical depth (i.e., toward higher layers of the solar atmosphere), ranging from approximately 240 G at $\log(\tau) = 0.0$ to nearly 115 G at $\log(\tau) = -1.0$. Beyond $\log(\tau) = -1.0$, it appears to stabilize, maintaining a nearly constant value around 100 G.

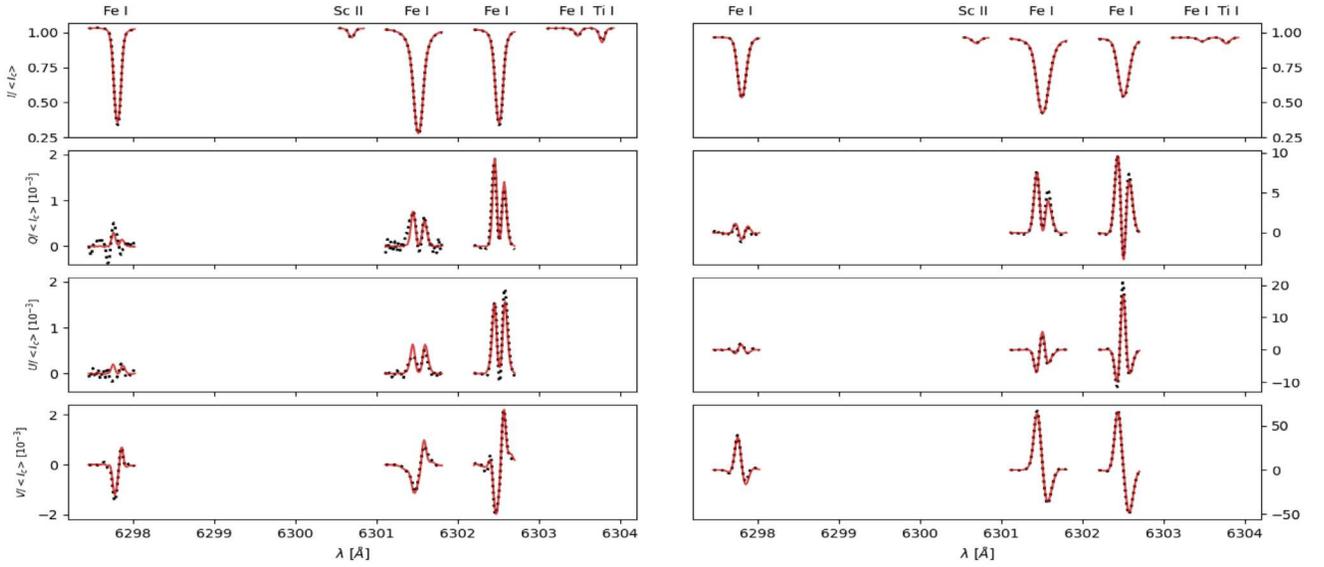

**Figure 8.** Examples of Stokes profiles observed by ViSP at DKIST telescope displayed as black dots. The red line represents the fit achieved in the inversion. Left and right panels stand for Stokes profiles observed in the internetwork and the network, respectively.

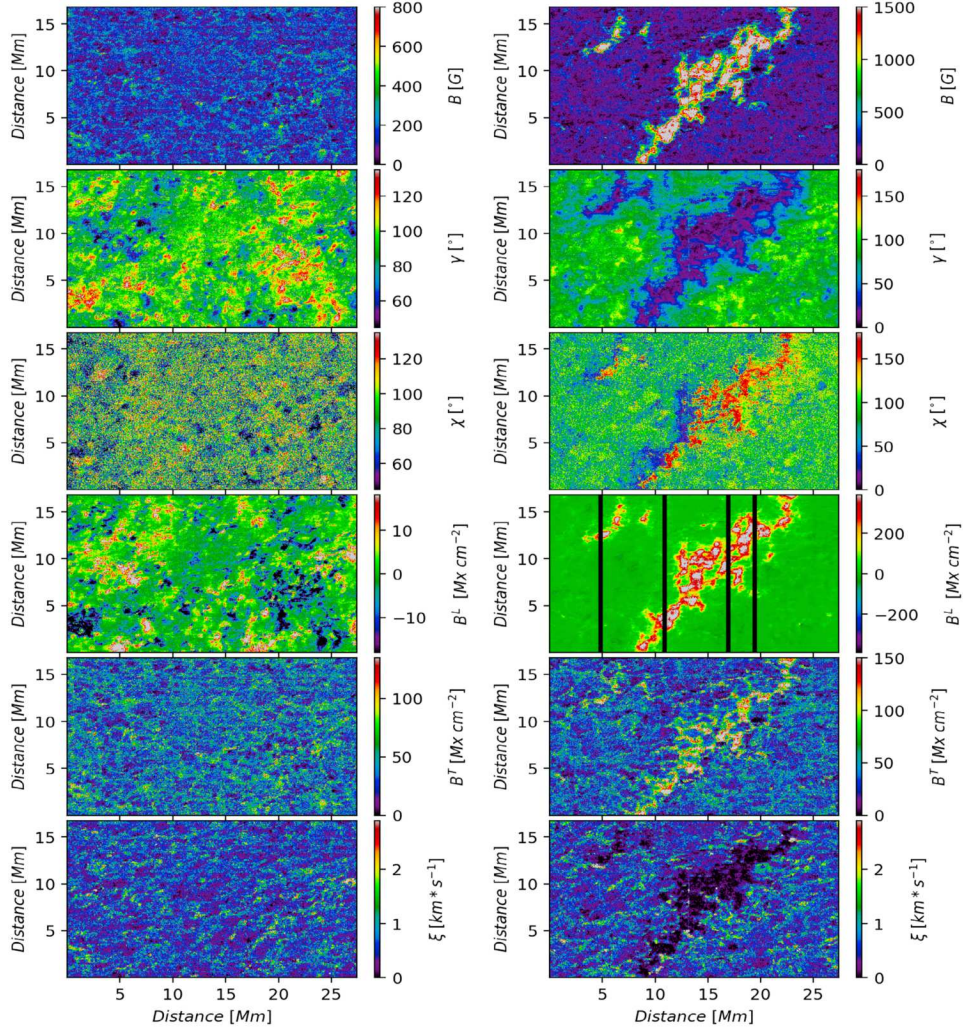

**Figure 9.** Results of the inversion applied to real ViSP observations. Surface maps (from top to bottom) of magnetic field strength, inclination, and azimuth, longitudinal and transverse magnetic field, and microturbulence velocity at $\log(\tau_{500}) = -0.8$. The left and right panels represent the results for the internetwork and the network, respectively. The four black, vertical lines on $B^L$ network panel are the locations selected to make Figure 11.

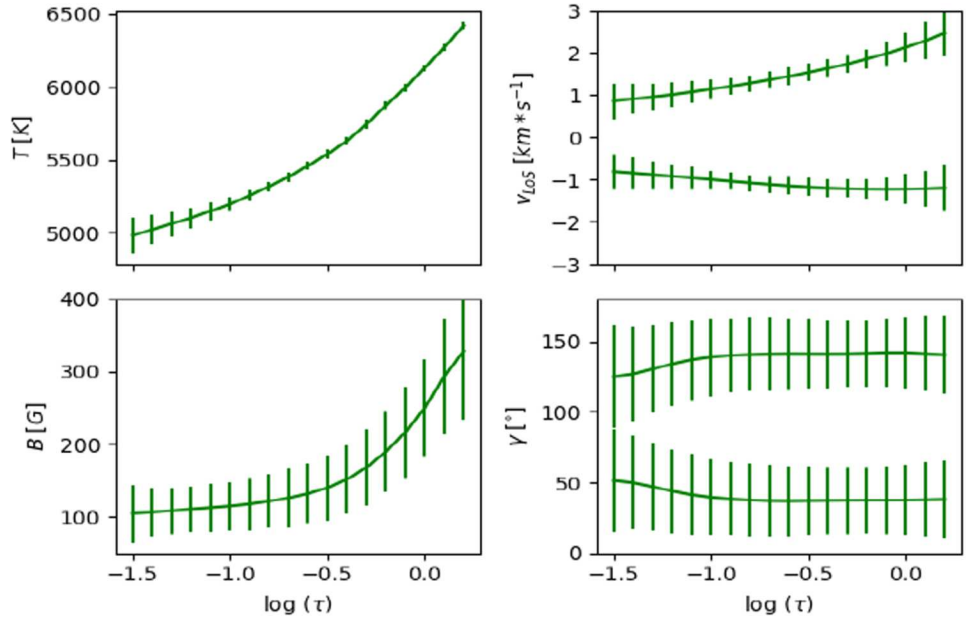

**Figure 10.** Results of the inversion applied to real ViSP observations as a function of the logarithm of the optical depth at a reference wavelength of 500 nm, $\log(\tau)$, to temperature $T$ (top left), LoS velocity $v_{\mathrm{LoS}}$ (top right), the magnetic field strength $B$ (bottom left), and magnetic field inclination with respect to the observer line of sight $\gamma$ (bottom right). The error bars represent the $\sigma$ value of the differences between the simulation and the inversion in each optical depth obtained in Section 3.2.3.

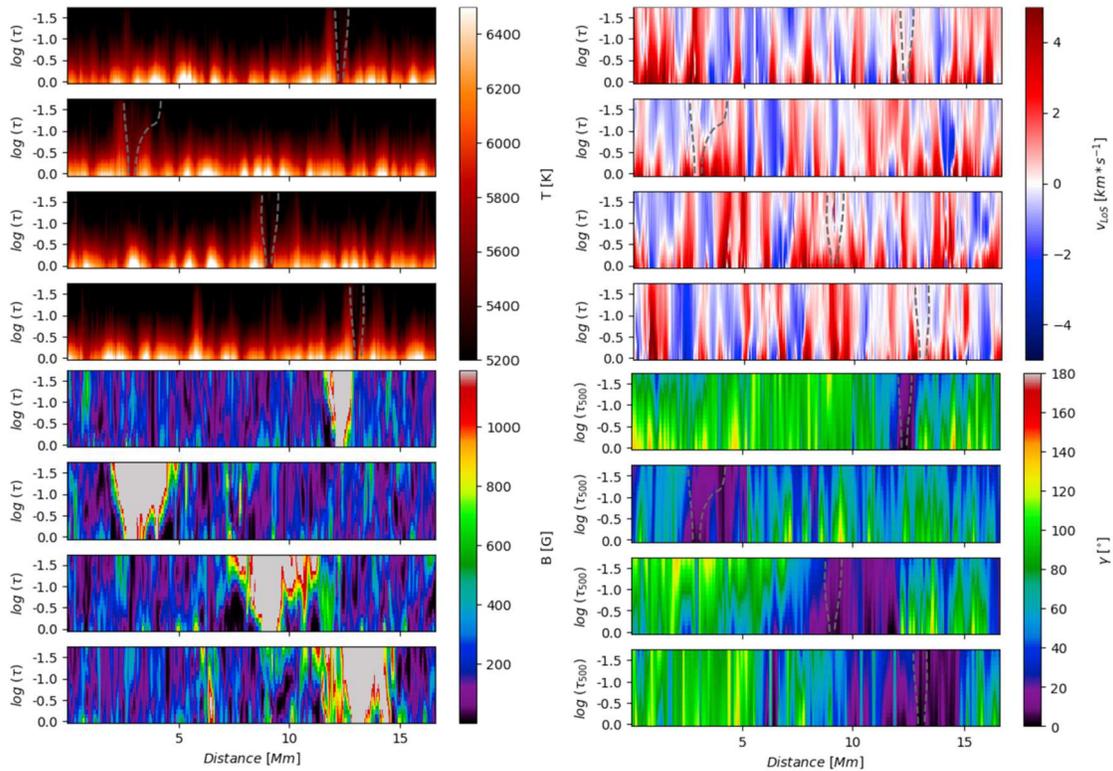

**Figure 11.** Results of the inversion applied to real ViSP observations. Depth maps of temperature $T$ (top left), LoS velocity $v_{\mathrm{LoS}}$ (top right), the magnetic field strength $B$ (bottom left), and magnetic field inclination with respect to the observer line of sight $\gamma$ (bottom right). In Figure 9 (Network $B^L$ panel) can be seen, in black color, the four locations selected for this plot. Gray dashed contours stand for magnetic fields above 1500 G.

The observed trend in the average magnetic field behavior aligns with findings from S. Danilovic et al. (2016), with minor discrepancies, particularly at $\log(\tau) = 0.0$, where a value of 130 G was reported. However, we argue that this discrepancy is not critical. As elucidated in J. C. Trelles Arjona et al. (2023), the average magnetic field strength in the internetwork is not a constant fixed value, but is contingent on long-term temporal variations and spatial locations across the solar surface. Consistently, our study, in line with previous research (D. Orozco Suárez et al. 2007; S. Danilovic et al. 2016; M. J. Martínez González et al. 2016; J. C. Trelles Arjona et al. 2021a), reaffirms that internetwork magnetism is composed

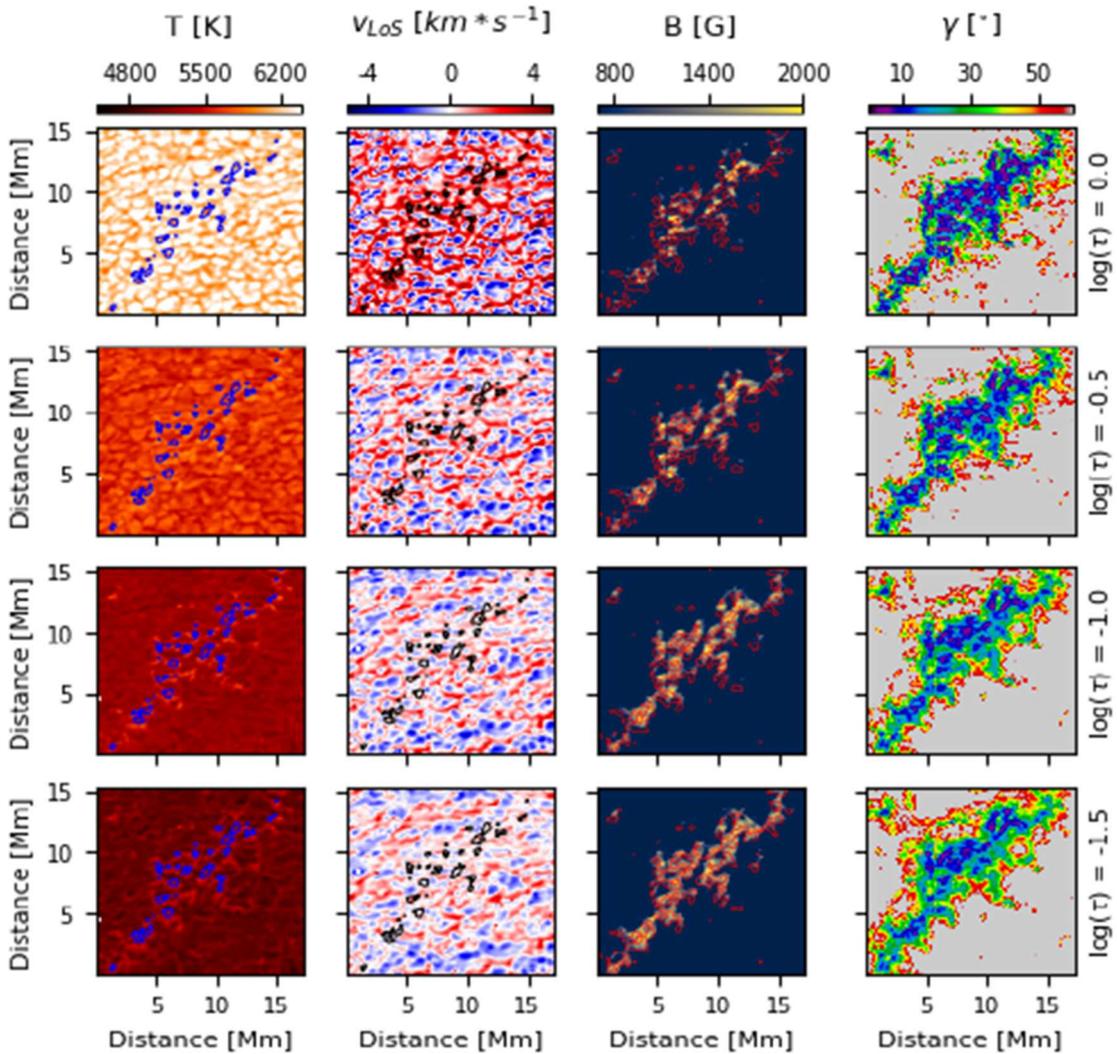

**Figure 12.** Results of the inversion applied to real ViSP observations for temperature $T$, LoS velocity $v_{\text{LoS}}$, the magnetic field strength $B$, and magnetic field inclination at different optical depths (from top to bottom, $\log(\tau) = 0.0$, $\log(\tau) = -0.5$, $\log(\tau) = -1.0$, and $\log(\tau) = -1.5$). Blue and black contours in temperature and LoS velocity maps represent magnetic field strengths over 1500 G at $\log(\tau) = -1.0$. Red contours in magnetic field strength maps stand for temperature above 5500 K at $\log(\tau) = -1.0$.

predominantly of hectogauss (hG) fields, some of them exceeding kG.

Figure 10 (bottom right) illustrates the results of the magnetic field inclination. In deeper layers, the fields remain relatively constant until $\log(\tau) = -1.0$, where a clear trend emerges, suggesting a transition of perpendicular magnetic fields toward a more horizontal orientation with increasing height. This observation corroborates the findings of S. Danilovic et al. (2016; see their Figure 11).

### 5.2. Thermodynamic and Magnetic Properties of Network

The high spatial resolution of the ViSP instrument at the DKIST telescope and multiline inversions allow us to revisit some long-standing debates in solar physics. J. O. Stenflo et al. (1984) found asymmetries in the Stokes $V$ profiles observed in the network. Subsequently, numerous studies have sought to uncover the thermodynamic and magnetic characteristics of the network flux tubes by interpreting these Stokes $V$ asymmetries (see U. Grossmann-Doerth et al. 1988; S. K. Solanki 1989; J. Sanchez Almeida et al. 1996; L. R. Bellot Rubio et al. 2000, and references therein). Our results, depicted in Figures 11 and 12, showcase temperature, LoS velocity, magnetic field strength, and magnetic field inclination for an entire network patch at four optical depths and selected vertical cuts, respectively. Although our findings align with the general understanding presented in prior work concerning flux tubes, we highlight certain distinctive results. In the temperature results, we detect a slightly lower temperature at the center of the flux tube compared to the quiet photosphere. We also identify temperature enhancements in specific locations near the flux tubes (see the left column of Figure 12 at high layers), consistent with the observations of U. Grossmann-Doerth et al. (1988) and C. U. Keller et al. (1990). This phenomenon is likely due to the horizontal component of the radiative flux, as proposed by P. Fabiani Bendicho et al. (1992).

Interestingly, this observed "hot cloud" of plasma is not situated at the central axis of the flux tubes but is notably offset to the tube walls. Furthermore, the distribution of the hot plasma cloud is nonuniform, with its location varying depending on the specific flux tube to which it belongs. This variability is likely due to the relatively rapid evolution of these structures and the surrounding environment. Compared with the quiet photosphere, U. Grossmann-Doerth et al. (1988) reported a 1000 K difference, while we observed approximately 500 K within the optical range of $\log(\tau) = -1.0$ to $\log(\tau) = -2.0$. This difference decreases beyond this optical range on both sides. In the top panel of Figure 12, the LoS velocity reveals pronounced downflows, particularly around the flux tube, reaching velocities exceeding $4\,\mathrm{km\,s^{-1}}$ in deep layers. This observation aligns with the findings of L. R. Bellot Rubio et al. (2000). Moreover, looking at the flux tubes in Figure 11 we detect nonnegligible mass motions within the magnetic elements. Above the canopy and inside the flux tubes, we find weak downflows and upflows. In L. R. Bellot Rubio et al. (2000) they conclude that mass motions within flux tubes intensify with depth, leading to larger zero-crossing shifts in spectral lines formed at lower atmospheric layers. The identification of downflows and upflows in the upper layers of flux tubes does not contradict their findings. In the absence of adequate spatial resolution, the opposing average velocities could counterbalance each other, misleadingly suggesting a greater drop in velocity in the upper layers than is actually the case.

The plots of Figures 11 and 12 referring to the magnetic field strength illustrate the presence of the canopy. The base of the flux tubes exhibits thinness and gradually expands with height. Notably, beneath the canopy, areas with nearly complete nonmagnetization are observed. The orientation of the magnetic field inclination in the flux tubes at their base is nearly perpendicular to the solar atmosphere, while the magnetic field within the flux tube progressively inclines toward a more horizontal orientation with increasing height. No discernible changes in the polarity of magnetic fields along the line of sight within the flux tubes are detected in our observations. In M. J. Martínez González et al. (2012), certain locations within the studied feature demonstrate a negative asymmetry in area (and amplitude) of Stokes $V$ profiles, implying a potential change in either the magnetic field or velocity field gradient. We propose that the negative area (and amplitude) in Stokes $V$ arises from a sign change in the velocity field gradient, as no sign changes in the longitudinal magnetic field are detected.

## 6. Conclusions

The data acquired by the ViSP instrument at the DKIST telescope represent a significant leap in both quality and spatial resolution. Employing multiline inversions proves essential for enhancing accuracy in determining the stratification of thermodynamic and magnetic properties of the solar atmosphere. Multiline inversions require the careful characterization of parameters of the spectral lines to be used. With this purpose we have determined the $\log(gf)$ values of the spectral lines around 630.1 nm, typically observed with one of the arms of the ViSP instrument at DKIST telescope. To find the optimal inversion strategy that involves the spatial resolution and spectral lines in this arm of the ViSP instrument, we have relied on MHD simulations. We have found that, in this spectral range, relying solely on intensity profiles in inversions is less effective than at 1.5 $\mu$m, with the difference becoming even more pronounced in deeper layers of the solar atmosphere. Several factors contribute to these differences between spectral ranges. One key aspect is the difference in the height of formation of the spectral lines, both in absolute terms and in the range of heights they sample (J. M. Borrero et al. 2016). Additionally, the infrared spectral range exhibits greater variations in excitation potential, $\log(gf)$ values, and Landé factors among its lines compared to the visible range. For multiline inversions, an optimal selection of spectral lines should ideally include a broad range of excitation potentials, $\log(gf)$ values, and effective Landé factors. This diversity improves the capacity to accurately retrieve atmospheric structure and magnetic properties. Regarding magnetic field measurements, incorporating a spectral line with a Landé factor of 0.0 could be particularly valuable if possible, as it would further increase the disparity in sensitivity to the magnetic field across different lines, enhancing our ability to disentangle thermodynamic and magnetic properties. We also found that by increasing the number of spectral lines in the inversions, the results systematically improve. We have shown that the best strategy to study the internetwork magnetism in this spectral range of data acquired with the ViSP instrument is to use two runs of inversions. First, we used only intensity profiles of six spectral lines, and second, we incorporated the polarimetry of three spectral lines (those more intense in the spectral range). This inversion strategy enables the retrieval of the overall depth-dependent behavior present in the original simulation. Its application to real ViSP observations unveils insights into both internetwork and network properties. Notably, we observe a decrease with height in the averaged magnetic field strength within the internetwork at deeper layers, followed by a relatively constant trend at higher layers. Furthermore, our analysis reveals the constancy of perpendicular fields up to $\log(\tau) = -1.0$, beyond which they transition to a more horizontal orientation. The analysis presented here about the network confirms the global picture shown in previous studies. We show the depth dependence of the network canopy. We distinguish upflows and downflows inside the flux tubes. In addition, we encounter an enhancement of temperature in particular areas around the network, with a difference of more than 500 K compared to the quiet photosphere. Although these findings contribute valuable information, further research is imperative for a robust interpretation. Exploring multiline, multiarm inversions of ViSP data to enhance the accuracy of the results and encompass additional atmospheric layers emerges as the next logical step.

## Acknowledgments

The author is especially grateful to Dr. Basilio Ruiz Cobo and Dr. Valentín Martínez Pillet for their valuable support, insightful discussions, and helpful comments that contributed to improving the manuscript. The author acknowledges financial support from the University of Colorado Boulder through the George Ellery Hale fellowship. The observations used in this study were taken with the DKIST telescope, located at Haleakala Observatory on the Hawaiian island of Maui (USA). The research reported herein is based in part on data collected with the Daniel K. Inouye Solar Telescope (DKIST), a facility of the National Solar Observatory (NSO). NSO is managed by the Association of Universities for Research in Astronomy, Inc., and is funded by the National Science Foundation. DKIST is located on land of spiritual and cultural significance to Native Hawaiian people. The use of this important site to further scientific knowledge is done so with appreciation and respect. This paper made use of the IAC Supercomputing facility HTCondor (http://research.cs.wisc.edu/htcondor/), partly financed by the Ministry of Economy

## Appendix

One of the distinct advantages of performing multiline inversions is the enhanced spatial resolution in depth. To evaluate the improvement as additional lines are introduced, it is valuable to compare the simulation and the results of the inversion across each atmospheric layer.

In this appendix, we present the Pearson correlation coefficients between the simulation and the results of the inversion for each optical depth, allowing us to assess the depth-dependent evolution of all experiments with simulations. Figures 13–16 display these coefficients for the different inversion setups discussed in the main text. The background colors green, yellow, and red indicate good, moderate, and poor correlations, respectively. Hence, values closer to unity are preferable. The correlation value $r$ in each plot represents the average across the entire optical depth range (between $\log(\tau) = 0.5$ and $\log(\tau) = -1.5$).

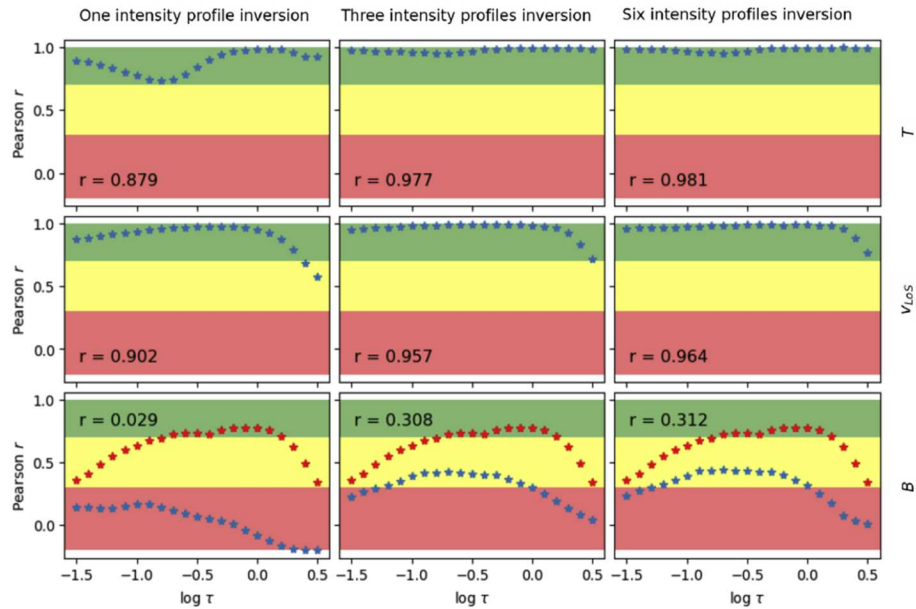

**Figure 13.** Pearson's correlation coefficient between some parameters (from top to bottom: temperature, LoS velocity, and magnetic field strength) in the original simulation, and the same parameters inferred by the inversion of one, three, and six intensity profiles (first, second, and third columns from the left, respectively). Blue stars stand for the Pearson's correlation coefficient derived in this work. Red stars stand for the Pearson's correlation coefficient derived in J. C. Trelles Arjona et al. (2021a). The values closer to unity are better.

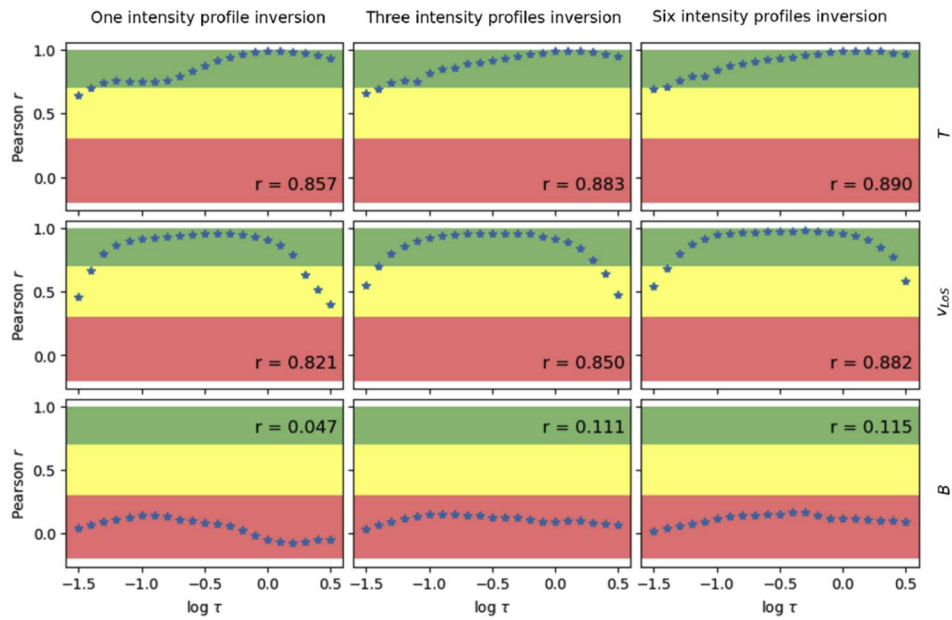

**Figure 14.** Pearson's correlation coefficient between some parameters (from top to bottom: temperature, LoS velocity, and magnetic field strength) in the original simulation, and the same parameters inferred by the inversion of one, three, and six intensity profiles (first, second, and third columns from the left, respectively). The values closer to unity are better.

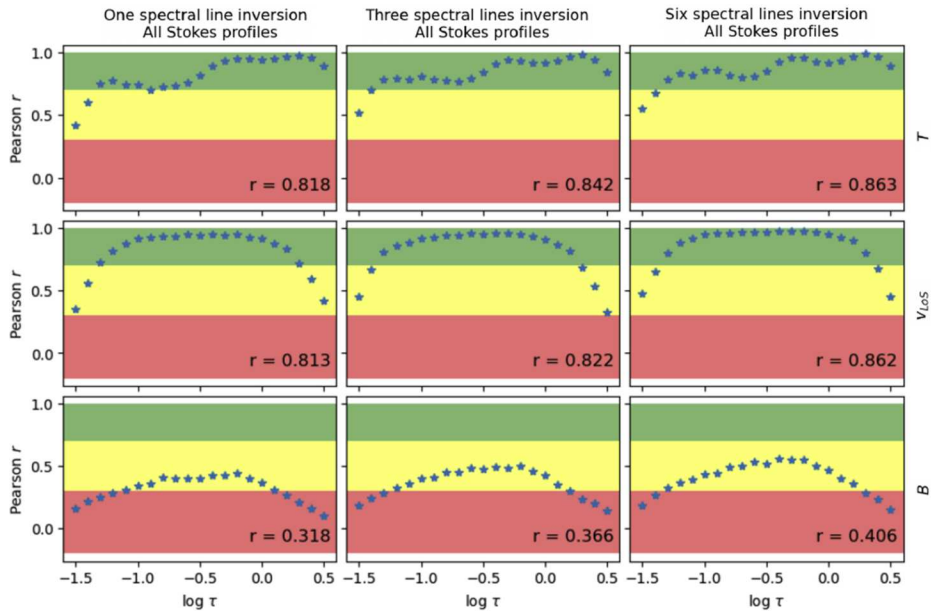

**Figure 15.** Pearson's correlation coefficient between some parameters (from top to bottom: temperature, LoS velocity, and magnetic field strength) in the original simulation, and the same parameters inferred by the inversion of one, three, and six full Stokes profiles ($I$, $Q$, $U$, and $V$) (first, second, and third columns from the left, respectively). The values closer to unity are better.

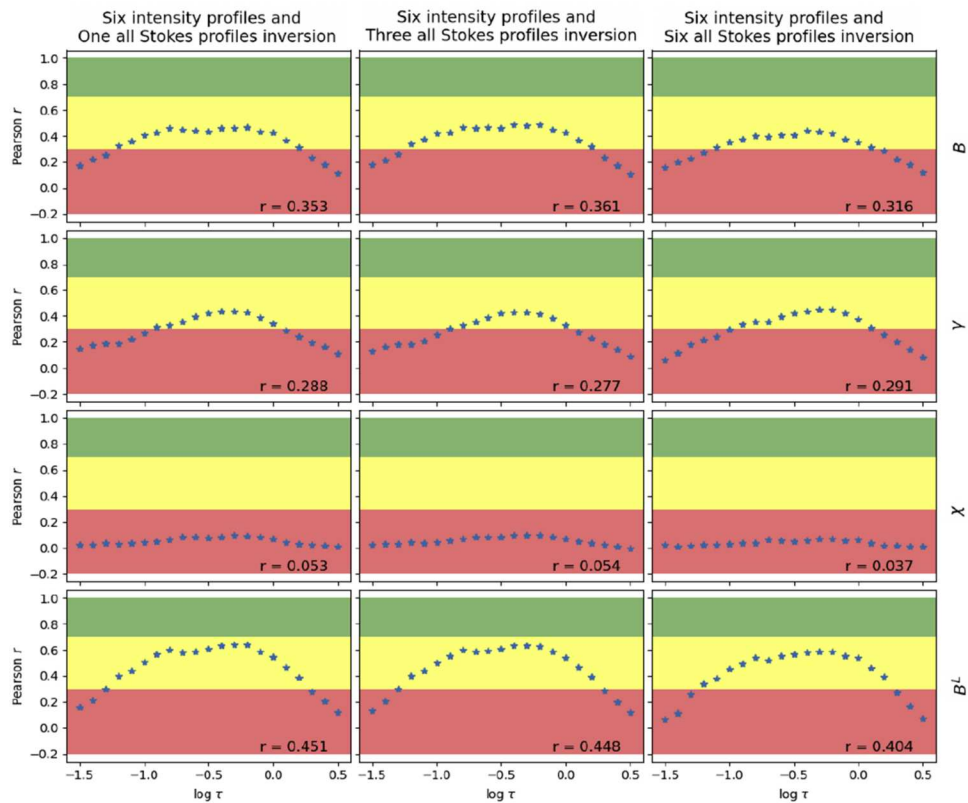

**Figure 16.** Pearson's correlation coefficient between some parameters (from top to bottom: magnetic field strength, inclination, and azimuth and longitudinal magnetic field) in the original simulation, and the same parameters inferred by the inversion of six spectral lines in intensity and one, three, and six full Stokes profiles (I, Q, U, and V; first, second, and third columns from the left, respectively). The values closer to unity are better.

## ORCID iDs

J. C. Trelles Arjona  https://orcid.org/0000-0001-9857-2573